\documentclass[twocolumn]{aastex62}

\usepackage{amsmath} 
\usepackage{graphicx} 
\usepackage{textcomp}
\usepackage{gensymb}
\usepackage{natbib} 
\usepackage{longtable}
\usepackage{multirow}
\usepackage{float}

\received{29 March 2019}
\revised{1 August 2019}
\accepted{16 August 2019}

\shorttitle{Variability in S and VV Corona Australis}
\shortauthors{Sullivan et al.}

\begin{document}
\title{S and VV Corona Australis: Spectroscopic Variability in Two Young Binary Star Systems}
\author[0000-0001-6873-8501]{Kendall Sullivan}
\affil{University of Texas at Austin, Austin TX 78712, USA}
\affil{Lowell Observatory, Flagstaff AZ 86001, USA}

\author{L. Prato}
\affil{Lowell Observatory, Flagstaff AZ 86001, USA}

\author{Suzan Edwards}
\affil{Smith College, Northampton MA 01063, USA}

\author{Ian Avilez}
\affil{Lowell Observatory, Flagstaff AZ 86001, USA}

\author{Gail H. Schaefer}
\affil{The CHARA Array of Georgia State University, Mount Wilson Observatory, Mount Wilson CA 91023, USA}

\begin{abstract}
We used high-resolution near-infrared spectroscopy from the NIRSPEC instrument on the Keck II telescope, taken over multiple epochs spanning five years, to examine two young binary T Tauri star systems, S Corona Australis and VV Corona Australis. The stars in these 1-2" separation systems have optically thick circumstellar disks and high extinctions at optical and near-infrared wavelengths. Using a combination of new and archival data, we have determined the spectral types of all the stars in these two systems for the first time, examined the variable NIR veiling, measured the emission line equivalent widths, and created spectral energy distributions. They have similar spectral types (K7-M1) and are at approximately the same evolutionary stage, allowing comparison of the four stars in the two systems. We conclude that S CrA and VV CrA are young binary systems of stars bridging the Class I and Class II evolutionary stages, characterized by high accretion luminosities and variable emission lines.
\end{abstract}

\clearpage
\section{Introduction}
Young stars with circumstellar disks provide the opportunity to examine the environments supporting the earliest stages of planet formation. Most stars form in binary or higher-order multiple systems \citep[e.g.,][]{1991A&A...248..485D, 2010ApJS..190....1R}, dynamic environments with the potential to significantly modify circumstellar disks \citep[e.g.,][]{1996ApJ...458..312J, 2012ApJ...745...19K, 2005ApJ...619L.175A}. Young star evolution, circumstellar disk evolution, and thus planet formation are subject to star-disk interactions in binaries. The frequency, distribution, and masses of planets may be inhibited by these interactions \citep{2016AJ....152....8K}. Recent large scale theoretical and observational studies of young binary stars \citep[e.g.,][]{2017MNRAS.468.1631R, 2018MNRAS.475.5618B, 2019ApJ...872..158A} have emphasized the need for greater understanding of these complex systems on both population-wide and system-specific scales.

In order to understand the likelihood of planet formation in circumstellar disks surrounding the stars in binary and higher-order multiple systems, we must understand young binary star systems themselves. Envelopes accrete onto disks and disks in turn accrete onto the central stars, triggering excess continuum emission, line emission, outflows, and stellar jets. By understanding the complex potential star-star and star-disk interactions at the earliest observable stages of stellar evolution, we may begin to draw a more complete picture of the processes that shape planet formation. Near-infrared (NIR) spectroscopy provides a particularly powerful tool for studying very young stars because NIR observations can penetrate much of the dust around young stars, revealing their photospheres, and spectroscopy provides information about both star and disk properties. The NIR is also the most efficient wavelength regime in which to study low-mass ($< 1 M_{\odot}$) young stars, given that their emission peaks in the NIR. Multi-epoch spectroscopy furthermore yields insight into the nature of young star variability.

\begin{deluxetable*}{cccccccc}
\tablecaption{System Parameters and 2MASS/WISE JHKL Magnitudes for S CrA and VV CrA \label{table:mags}}
\tablecolumns{8}
\tablewidth{0pt}
\tablehead{
\colhead{Star} & \colhead{Separation} & \colhead{Separation} & \colhead{Position} & \colhead{J} & \colhead{H} & \colhead{K} & \colhead{L}\\
\colhead{} & \colhead{(arcsec)\tablenotemark{a}} & \colhead{(AU)\tablenotemark{b}} & \colhead{Angle ($^{\circ}$)\tablenotemark{a}} & \colhead{(mag)} & \colhead{(mag)} & \colhead{(mag)} & \colhead{(mag)}
}
\startdata
S CrA & 1.3$"$ & 195 & 160 & 8.19$\pm$0.02 & 7.05$\pm$0.02 & 6.11$\pm$0.02 & 5.14$\pm$0.01\\
VV CrA & 1.9$"$ & 285 & 45 & 9.86$\pm$0.03 & 7.86$\pm$0.04 & 6.24$\pm$0.02 & 4.38$\pm$0.01\\ 
\enddata
\tablenotetext{a}{From \citet{2003ApJ...584..853P}.}

\tablenotetext{b}{Assuming a distance of 150 pc.}

\end{deluxetable*}

S Corona Australis (S CrA) and VV Corona Australis (VV CrA) are two young binary systems, both located in the southern star forming region Corona Australis, which is located $\sim$150 pc away \citep[the Gaia parallax for VV CrA is $6.7 \pm$ 0.1 milliarcseconds;][]{2018A&A...616A...9L,2018A&A...616A...1G, 2016A&A...595A...1G} and has an estimated age of 1-3 Myr \citep{1998AJ....115.1617C, 1997ApJ...478..295C}. System separations and unresolved JHKL magnitudes from 2MASS and WISE are shown in Table \ref{table:mags}. The stars in these systems are all in the T Tauri phase \citep{1945ApJ...102..168J}, characterized by variability and prominent hydrogen emission lines produced by accretion. Given the variability of these systems, we define the primary and secondary stars using our flux ratio measurements in 2015, such that S CrA A is the NW component, and VV CrA A is the southern component. S CrA and VV CrA are young enough that their photospheres remain heavily obscured \citep[$A_{v} \sim 2$ for both components of S CrA and VV CrA A, and $A_{v} = 10.2$ for VV CrA B, ][]{2018A&A...617A..83V}, although the systems themselves are bright in the NIR, the result of excess emission from their warm, optically thick circumstellar disks and high accretion rates. Both S CrA and VV CrA exhibit variable optical and NIR veiling over time \citep{1980AJ.....85..555B, 2002A&A...384.1038E, 2005ESASP.560.1021W, 2008A&A...488..997K, 2012ApJS..201...11K, 2012A&A...543A.162V, 2016MNRAS.458.2476S}, indicating that they are both experiencing variable accretion. 

S CrA is one of the earliest stars identified as a T Tauri variable \citep{1945ApJ...102..168J} and was identified as an $\approx 1"$ binary in the same paper. Most studies of this system in the NIR have focused on understanding and characterizing its accretion and veiling \citep[e.g.,][]{2005ESASP.560.1021W, 2008A&A...482L..35G, 2012AstL...38..649D, 2014A&A...568L..10P}. \citet{2019arXiv190402409C} recently observed the disks of the two stars in this binary directly with ALMA, while other groups have used NIR and MIR interferometry to study the disks of the two stars in the binary \citep{2009A&A...502..367S, 2014A&A...568L..10P, 2017A&A...608A..78G}. Notably, \citet{2013A&A...551A..34S} found that the SED of S CrA suggests that at least one component has a very massive, flared disk, and that the binary may be surrounded by cloud or envelope material. Other observations have focused on specific lines to examine disk and accretion dynamics, primarily in the IR \citep{2007A&A...464..687C, 2013A&A...559A..77F}. \citet{2003ApJ...584..853P} used R $\approx$ 800 NIR spectroscopy to observe the photospheres of S CrA A and B, determine the K-band veiling, and estimate spectral types for the stars; however, these results were relatively uncertain given the low spectral resolution. Gaia DR2 parallaxes \citep{2018AJ....156...58B} to visual binaries such as S CrA are known to be unreliable, so we assume the same 150 pc Gaia distance for S CrA as was found for VV CrA, which is not a visual binary.

Most previous work on VV CrA has treated it as a single star rather than a binary system. The secondary component is an IR companion \citep[IRC,][]{1997ApJ...480..741K}, a PMS star with a red SED, and is extremely faint in optical wavelengths; even in the NIR, it was not identified as a binary until 1992 \citep{1992PASP..104..479G}. VV CrA has primarily been studied in two capacities: first, as an IRC host system \citep{1997ApJ...480..741K}, and second, with spectroscopy to determine disk composition \citep[e.g.,][]{2000A&AS..144..285Y, 2009ApJ...696L..84C, 2011ApJ...729..145K, 2015A&A...584A..28L}. \citet{2016MNRAS.458.2476S} produced detailed modeling of both components of VV CrA and their disks based on data taken at 1.3 mm, in the mid infrared, and in the K-band. They were not able to determine the spectral types of the stars in the system because strong veiling obscured the photospheric absorption lines at the time of their K-band observations.

In contrast to these previous studies, we have been able to observe the photospheres of all components of S CrA and VV CrA in multiple epochs of high-resolution NIR spectra. We used high-resolution J-, H-, and K-band spectroscopy taken with the NIRSPEC instrument on the Keck II telescope to determine stellar spectral types, quantify the H-band veiling, and characterize the hydrogen emission lines. We estimated accretion luminosity, searched for spectral signatures of accretion, winds, and jets, and constructed spectral energy distributions (SEDs) using both new and archival data. We describe our data and observations in Section \ref{obs}, present our analysis and results in Section \ref{anr}, discuss our results in Section \ref{disc}, and summarize in Section \ref{sum}.

\section{Observations}\label{obs}
\subsection{Spectroscopy}

\begin{deluxetable}{ccc}
\tablecaption{Keck/NIRSPEC observations \label{table:observations}}
\tablecolumns{3}
\tablewidth{0pt}
\tablehead{
\colhead{UT} & \colhead{Band} & \colhead{Target} \\
\colhead{Date} & \colhead{} & \colhead{Star} 
}
\startdata
2002/06/23 & H & VV CrA \\
2002/07/18 & H & S CrA \\
2003/05/14 & H & VV CrA \\
2003/09/07 & H & VV CrA \\
2003/09/07 & K & VV CrA \\
2004/07/13 & H & VV CrA \\
2004/07/22 & H & S CrA, VV CrA \\
2007/04/30 & H & S CrA \\
2007/08/09 & H & S CrA \\
2008/05/23\tablenotemark{a} & J & S CrA, VV CrA \\
2008/05/23\tablenotemark{a} & K & S CrA, VV CrA \\
\enddata

\tablenotetext{a}{Archival data (PI G. Herczeg).}

\end{deluxetable}

\begin{figure*}
\includegraphics[width = \linewidth, keepaspectratio = true]{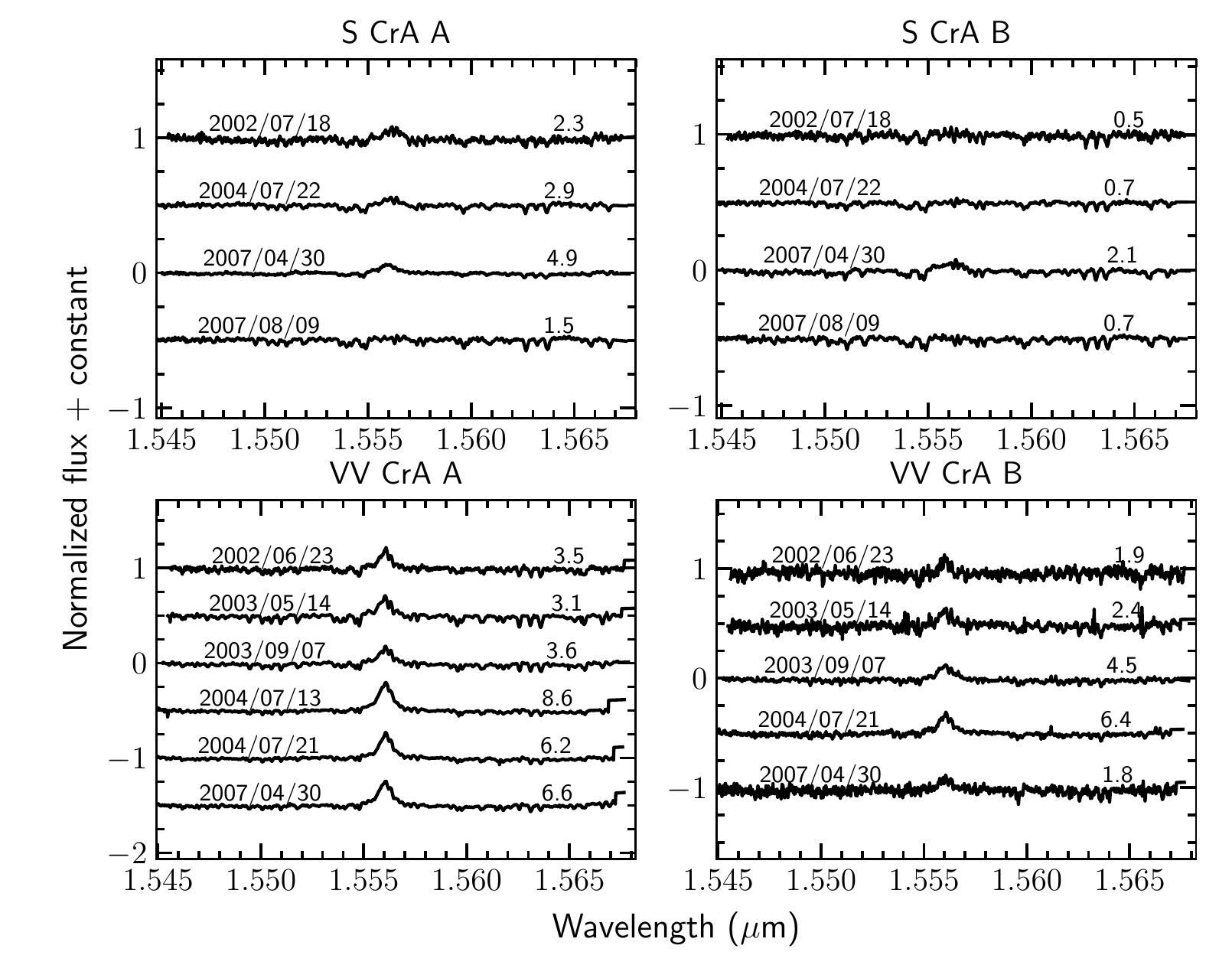}
 \caption{NIRSPEC order 49 H-band spectra of target binaries. Both Brackett 16 emission and numerous atomic and molecular absorption features are present in this order; particularly useful are the two FeI lines flanking an unresolved OH doublet at $\sim$1.563 $\mu$m. The variations in absorption line depths are caused by variable veiling. The UT dates of observation and veiling values are indicated on the left and right side of the figure, respectively.} 
\label{fig:all spectra}
\end{figure*}

\begin{figure*}
\includegraphics[width = \linewidth, keepaspectratio = true]{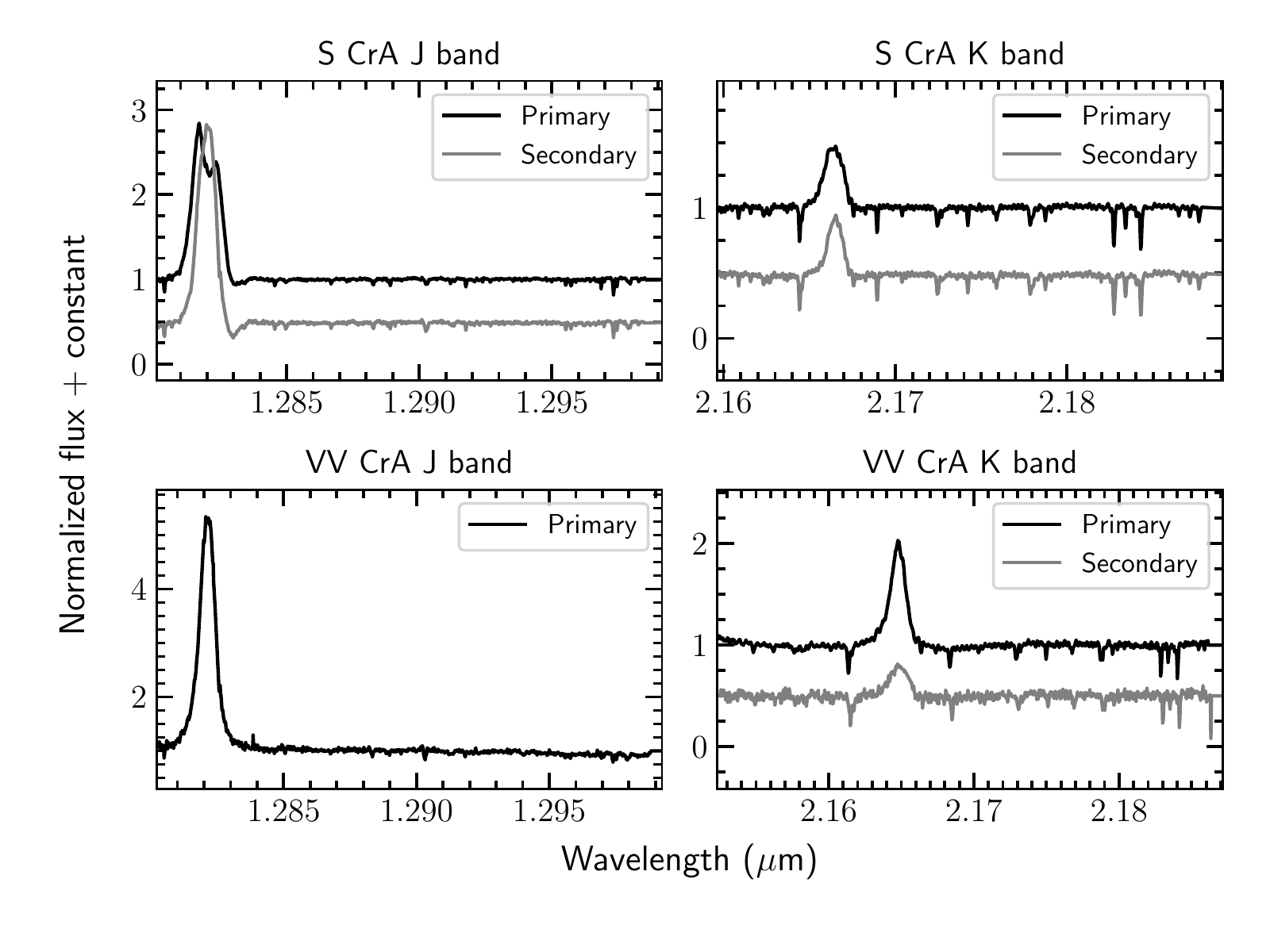}
 \caption{J- (order 59, left column) and K-band (order 35, right column) spectra for S CrA and VV CrA showing the Paschen $\beta$ and Brackett $\gamma$ emission lines, respectively. Only the primary component of VV CrA was detectable in the J band at the time of observation. The data were all taken on UT 2008 May 23. These spectra have not been divided by a telluric standard star as they were only used for measuring emission line strengths. The components are offset by an arbitrary constant for presentation purposes.} 
\label{fig:JK spectra}
\end{figure*}

We observed S CrA and VV CrA in the H-band (central wavelength = $1.555 \mu$m) at the Keck II telescope on Mauna Kea with NIRSPEC, a cross-dispersed, cryogenic, NIR spectrograph employing a 1024x1024 ALADDIN InSb array detector. In addition to our own multi-epoch H-band observations of S CrA and VV CrA spanning from 2002 to 2007 (Figure \ref{fig:all spectra}), we used a single epoch of archival observations in the J and K bands, taken by G. Herczeg at Keck II using NIRSPEC in 2008 (Figure \ref{fig:JK spectra}). A summary of our observations, including observation dates and other details, is presented in Table \ref{table:observations}. For all observations, the slit size was 0.288$\times$24$''$, except for the J-band on UT 2008 May 23 when it was 0.288$\times$12$''$, yielding a resolution of $\sim$30,000 for all observations. For efficiency, we typically rotated the slit to observe both stars in each system simultaneously. 

The archival data were downloaded from the Keck Observatory Archive and we reduced both our own observations and the archival data identically using REDSPEC, a reduction software written for NIRSPEC data\footnote{http://www2.keck.hawaii.edu/inst/nirspec/redspec/index.html} \citep{REDSPEC}. The primary and secondary components of S CrA were observable in all bands, J, H, and K. Both components of VV CrA were detected in the H- and K-bands, but the secondary was much fainter than the primary star in J during the observations, therefore only VV CrA A was observed in J band.

Our data reduction procedure is described in \citet{2017ApJ...844..168K}. Briefly, to eliminate pixel to pixel inconsistencies, the data were pair subtracted and divided by a dark-subtracted flat field. REDSPEC performs both a spatial and a spectral rectification. The spectra were extracted, averaged, flattened, normalized, and corrected for barycentric motion. In the H-band we focused on NIRSPEC order 49 ($\lambda = 1.545 -1.570 \mu $m, Figure \ref{fig:all spectra}), which is useful because it lacks telluric lines (removing the need to divide by a telluric standard star) and has absorption features that allow spectral typing of low-mass stars. The archival J- and K-band data and our own K-band data from 2003 were processed in the same manner (Figure \ref{fig:JK spectra}). Because we were primarily interested in measurements of hydrogen emission lines in the J- and K-bands, we extracted order 59 (J-band) and order 35 (K-band), which both have prominent hydrogen emission lines, and we did not divide by a telluric standard star.

\subsection{Photometry}
\begin{deluxetable}{cCC}
\tablecaption{Flux Ratios and Binary Properties Observed on UT 2015 July 12. \label{table:flux_ratios}}
\tablecolumns{3}
\tablewidth{0pt}
\tablehead{
\colhead{} & \colhead{S CrA} & \colhead{VV CrA}
}
\startdata
Jcont & 0.430 $\pm$ 0.015 & 0.0152 $\pm$ 0.0012\\
Hcont & 0.373 $\pm$ 0.023 & 0.0981 $\pm$ 0.0050\\
Kcont & 0.333 $\pm$ 0.012 & 0.5207 $\pm$ 0.0096\\
Lp & 0.3421 $\pm$ 0.0022 & 1.01 $\pm$ 0.26 \\
$\rho ('')$ & 1.3093 $\pm$ 0.0013 & 2.1098 $\pm$ 0.0016 \\
P.A. ($^\circ$) & 153.021 $\pm$ 0.059 & 44.920 $\pm$ 0.048\\
JY & 2015.9313 & 2015.9313\\
\enddata
\end{deluxetable}

We present JHKL flux ratios between the primary and secondary for S CrA and VV CrA in Table \ref{table:flux_ratios}. We imaged the target systems using the NIRC2 adaptive optics camera at Keck II \citep{2000SPIE.4007....2W} on UT 2015 July 12. The images were reduced following the methods described in \citet{2014AJ....147..157S, 2018AJ....155..109S}. We measured the position angle and flux ratio of the secondary relative to the primary through PSF fitting, using the primary star as a PSF. We corrected the binary positions using the geometric distortion solution published by \citet{2016PASP..128i5004S}. 

For each system, Table \ref{table:flux_ratios} shows the secondary/primary flux ratio, the binary separation ($\rho$), position angle measured east of north (P.A.), and the Julian year (JY) of the observation. Compared to the photometry published in \citet{2003ApJ...584..853P}, it is clear that significant changes have occurred in the component flux ratios. For example, in VV CrA, the J flux ratio changed from 0.25 in 1996 to 0.015 in 2015. Noticeable relative motion has also taken place in these binary target systems (Tables \ref{table:mags} and \ref{table:flux_ratios}): the position angle of VV CrA has been relatively constant although the separation has increased by about 0.2$''$. In contrast, the S CrA binary separation seems to be constant; however, the position angle has changed by about 7 degrees.

\section{Analysis and Results}\label{anr}
\subsection{Higher Order Multiplicity}
We cross-correlated the H-band spectra for each target against the highest signal-to-noise spectrum of the object and found no apparent radial velocity shift that would suggest higher-order multiplicity of any of the stars in S CrA and VV CrA. We inspected the $JHKL$ NIRC2 AO images and residuals from the PSF fitting to check for any visible companions. We found no evidence for additional components within 0$\farcs$1 to 6$\farcs$0 and $\Delta K < 5.2$ mag. For a companion in a 1-10 AU orbit with a 3$\sigma$ detection, we set an upper limit on the mass of 50 - 100 $M_{Jup}$, assuming velocity resolution of 1 km/s, requiring a detection of $\sim$ 3 km/s in RV for a 0.6 $M_{\odot}$ star.

 \subsection{Age Approximation}
To compare age estimates with stellar evolutionary stage, we used the \citet{baraffe} isochrones and component star J band absolute magnitudes derived from 2MASS with the system flux ratios in Table \ref{table:flux_ratios}. We converted our spectral types to temperatures using \citet{2003ApJ...590..348L}. We de-reddened our J-band magnitudes using the $A_{\lambda}$ relation in section 3.5 but did not correct for J-band veiling, as the lack of photospheric lines in J prevented us from measuring the veiling. However, excess emission from the disk is typically least significant in the J band \citep[e.g.,][]{2012PASP..124.1137F, 2006ApJ...638..314D}. To convert to absolute magnitude, we assumed that the two systems fall at the same distance, and used d$=$150 pc, derived from the Gaia parallax measurement of VV CrA, which should be uncontaminated by the IRC which has not been detected at optical wavelengths. This distance yielded system ages of 1-2 Myr on the \citet{baraffe} tracks, consistent with previous estimates for Corona Australis.

To within an uncertainty of 1$\sigma$, both components of S CrA fall on the 1 Myr isochrone, and the stars in VV CrA are both consistent with the 2 Myr isochrone. Given that the stars in VV CrA appear less evolved than those in S CrA, this distinction in ages may reflect the approximations in our estimates of absolute J magnitude and T$_{eff}$. The extremely high extinction at VV CrA B is likely local, and a result of its nature as an IRC, rather than an indicator of a large amount of undetected material dimming the VV CrA system as a whole. There is also ample evidence that young stars at apparently distinct evolutionary stages appear to have similar ages \citep[e.g.,][]{2012ApJ...750..125A}; although statistically clusters with lower disk fractions appear older than those with high disk fractions \citep{2008ApJ...686.1195H}, the particulars and timeline for circumstellar disk evolution and dissipation around individual stars are still not completely understood. 

\subsection{Spectral Type Determination and Veiling Measurement}
\begin{figure*}
\includegraphics[width = \linewidth, keepaspectratio = true]{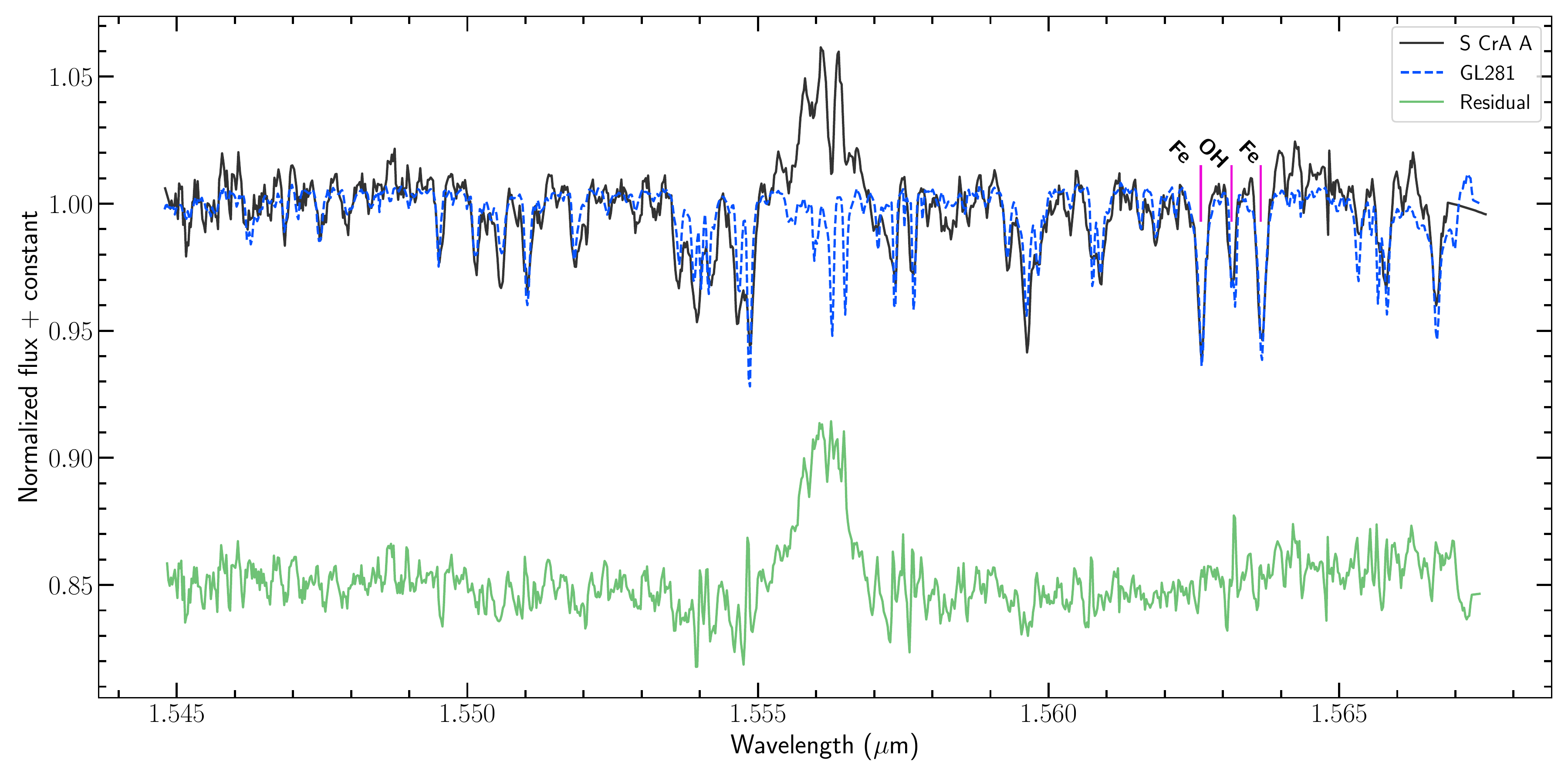}
 \caption{S CrA A (solid black) from 2004/07/22 UT, with the best fit template star, a K7 spectral type (dotted blue) scaled to the veiling value for that date (2.9) and overplotted, along with the residual spectrum from subtracting the data from the template (green). The Fe and OH lines used for spectral type determination are marked with vertical fuchsia lines. The K7 spectral template is GJ 281, which has a spectral type of K7V.} 
\label{fig:veil}
\end{figure*}

To measure spectral types and H-band veiling, $r_{H}$, we used templates from \citet{2002ApJ...569..863P} and \citet{2005AJ....129..402B}\footnote{The template spectra are available at \url{http://www2.lowell.edu/users/lprato/hband/homepage.html}}. We simultaneously matched the line ratios of the two Fe (1.562 and 1.563 $\mu$m) and one OH (1.5625$\mu$m) lines marked in Figure \ref{fig:veil} to measure the spectral type, and scaled our templates to match the veiling for each epoch. The best fit spectral type template for each star, along with the spectral type and luminosity class, are given in Table \ref{table:indices}. The measured $v \sin(i)$ of the S CrA components is 12 km/s \citep{2018A&A...614A.117G}, and the line widths of our spectral type template stars matched those of the the target stars well. Although we did not determine the $v \sin(i)$ of the VV CrA target stars, the line widths matched those of our template spectral type stars, which have e.g., a $v \sin(i) \sim 3.6$ km/s \citep{2010A&A...520A..79M} for the M0 star, GJ 763. Thus, we did not broaden the templates further to better match those of the data (e.g., Figure \ref{fig:veil}).

\begin{deluxetable*}{ccccCCCCC}
\tablecaption{Stellar properties \label{table:indices}}
\tablecolumns{9}
\tablewidth{0pt}
\tablehead{
\colhead{Star} & \colhead{Spectral Type} & \colhead{Best fit} & \colhead{Template}& \colhead{$T_{eff}$} & \colhead{$A_{V}$} & \colhead{$\alpha $} & \colhead{$M_{\star}/M_{\odot}$} & \colhead{$\log(L_{\star}/L_{\odot})$}\\
\colhead{} & \colhead{} & \colhead{template star} & \colhead{spectral type} & \colhead{(K)} & \colhead{(mag)} & \colhead{} & \colhead{} & \colhead{}
}
\startdata
S CrA A/NW & K7 & GJ 281 & K7V & 4000 & 2.0 & -0.17 &  0.7 & -0.18\\
S CrA B/SE & M1 & GJ 763/GJ 752a & M1V \tablenotemark{a}& 3700 &  2.0 & -0.17 & 0.45 & -0.45 \\
VV CrA A/S & M1 & GJ 763/GJ 752a & M1V \tablenotemark{a}& 3700 & 1.7 & 0.13 & 0.55 & -0.33\\
VV CrA B/N & M0 & GJ 763 & M0V &3850 & 10.2 & 0.37 & 0.45 & -0.45 \\
\enddata
\tablenotetext{a}{We did not have an M1 template, so the M1 spectrum is a hybrid spectrum made of an interpolation between M0 (GJ 763) and M2 (GJ 752a) templates.}
\tablecomments{The spectral index $\alpha$ is measured between 10.5 and 2.3 $\mu$m. The mass and luminosity were calculated assuming an age of 2 Myr using the \citet{baraffe} isochrones.}
\end{deluxetable*}

By comparing the line ratios in the $1.563 \mu$m spectral region where two FeI lines at 1.562 and 1.563 $\mu$m surround an unresolved OH doublet, we also determined a spectral type for each star. The OH/Fe line depth ratio is sensitive to temperature in low-mass stars. Thus, by comparing the observed line ratios between the Fe I and OH lines with a veiled accurately spectral typed template star we simultaneously determined the veiling and the spectral type. This spectral type determination is described further in \citet{2007ApJ...657..338P}, Figure 1. An example best fit veiled template spectrum of GJ 281, a K7 star, with a high SNR S CrA A spectrum is shown in Figure \ref{fig:veil}. In all, we used three different spectral type templates, a K7, M0, and M2 to measure our spectral types, and have summarized the template properties in Table \ref{table:indices}. We note that the observed line ratio for S CrA B and VV CrA A is well matched by an average between an M0 and an M2, which we treat as equivalent to an M1 spectral type. We estimate our spectral type uncertainty to be half of a spectral class, based on the clear difference in line ratios.

By doing this analysis for all epochs for each star, we determined component spectral types of these systems accurately for the first time: K7 for S CrA A, M1 for S CrA B, M1 for VV CrA A, and M0 for VV CrA B (Table \ref{table:indices}). Our templates are all main sequence stars, and the discrepancy in evolutionary stage between the templates and our target stars may introduce some error in our line ratios. However, tests of the divergence in OH/Fe line ratio with synthetic NextGen spectra \citep{1998A&A...337..403B} show that surface gravity is minimally important for stars with $T_{eff} < 4000$K.
 
\begin{deluxetable*}{c|CCC|CCC|CCC|CCC}
\tablecaption{Veiling and Brackett 16 emission equivalent widths \label{table:veiling}}
\rotate
\tablecolumns{5}
\tablewidth{0pt}
\tablehead{
\colhead{Date} & \multicolumn{3}{c}{S CrA A} & \multicolumn{3}{c}{S CrA B} & \multicolumn{3}{c}{VV CrA A} & \multicolumn{3}{c}{VV CrA B} \\
\colhead{} & \colhead{EW (\AA)\tablenotemark{a}} & \colhead{$r_{H}$} & \colhead{EW (\AA)\tablenotemark{b}} & \colhead{EW (\AA)\tablenotemark{a}} & \colhead{$r_{H}$} & \colhead{EW (\AA)\tablenotemark{b}} & \colhead{EW (\AA)\tablenotemark{a}} & \colhead{$r_{H}$} & \colhead{EW (\AA)\tablenotemark{b}} & \colhead{EW (\AA)\tablenotemark{a}} & \colhead{$r_{H}$} & \colhead{EW (\AA)\tablenotemark{b}} 
}
\startdata
2002/06/23 & \nodata & \nodata & \nodata & \nodata & \nodata & \nodata & 1.3 & 1.7 & 3.5 & 0.8& 1.3 & 1.9 \\
2002/07/18 & 0.7 & 2.2 & 2.3 & 0.2 & 1.2 & 0.5 & \nodata & \nodata & \nodata & \nodata & \nodata & \nodata \\
2003/05/14 & \nodata & \nodata & \nodata & \nodata & \nodata & \nodata & 1.4 &1.3 & 3.1 & 1.0 & 1.5 & 2.4 \\
2003/09/07 & \nodata & \nodata & \nodata & \nodata & \nodata & \nodata & 1.3 & 1.8 & 3.6 & 1.1 & 3.1 & 4.5 \\
2004/07/13 & \nodata & \nodata & \nodata & \nodata & \nodata & \nodata & 2.1 & 3.0 & 8.6 & \nodata & \nodata & \nodata\\
2004/07/22 & 0.5 & 4.2 & 2.9 & 0.2 & 2.3 & 0.7 & 1.8 & 2.4 & 6.2 & 1.3 & 4.0 & 6.4 \\
2007/04/30 & 0.7 & 6.0 & 4.9 & 0.7 & 2.1 & 2.1 & 1.9 & 2.5 & 6.6 & 0.8 & 1.3 & 1.2 \\
2007/08/09 & 0.3 & 3.5 & 1.5 & 0.3 & 1.6 & 0.7 & \nodata & \nodata & \nodata & \nodata & \nodata & \nodata\\
\hline
Mean & 0.55 & 3.98 & 2.9 & 0.35 & 1.78 & 1.0 & 1.63 & 2.12 & 5.27 & 1.0 & 2.24 & 3.28\\
1 $\sigma$ variance & 0.17 & 1.37 & 1.26 & 0.21 & 0.43 & 0.64 & 0.31 & 0.57 & 2.01 & 0.19 & 1.11 & 1.91\\
\enddata
\tablenotetext{a}{As observed.}
\tablenotetext{b}{Veiling corrected EW}
\end{deluxetable*}

Veiling, $r_{\lambda}$, is defined by the ratio of excess flux compared to the expected stellar photospheric flux \citep{1999A&A...352..517F} at a particular $\lambda$. In practice, we found veiling by adding a flat continuum to the unveiled spectral type templates and renormalizing until the depth of the absorption lines matched between the veiled template and target spectra. The 1.555$\mu$m veiling values for each epoch and each star are shown in Table \ref{table:veiling} and in Figure \ref{fig:all spectra}, and range from 0.5 (the minimum value observed in S CrA B) to 8.6 (the maximum value observed in VV CrA A). Our veiling uncertainties are on the order of 0.1, which we found by slightly varying our code input for the veiling value and assessing the corresponding change in the output, as in Section 3.4, below regarding our Br16 equivalent widths. We focus on the veiling results from a single order (NIRSPEC order 49) of our H-band observations only. We did not have J and K band spectral type templates with which to measure the veiling at those wavelengths, and, as veiling is wavelength-dependent, J- and K-band veiling may differ from $r_{H}$.

\subsection{Emission Line Equivalent Widths}
We measured the equivalent widths of the Brackett 16 hydrogen emission line (Br16, $\lambda$ =1.556$\mu$m) for each star at every epoch. Because this line is adjacent to the spectral features used to determine veiling, we were able to correct the measured equivalent widths for the veiling at each epoch, such that $EW_{corrected} = EW_{observed} * (1 + r_{1.55})$. Our results are shown in Table \ref{table:veiling}: the veiling-corrected EWs are significantly larger than the observed EWs. We estimate the error in our equivalent width measurements to be approximately 0.1 \AA, based on the repeated measurements of the EW of the same emission line using slightly different parameters (continuum value, line start and end points, etc). The continuum determination and the signal to noise ratio of the spectra contributed most of the EW uncertainty.

We also measured the equivalent width of the Brackett $\gamma$ (Br$\gamma$, $\lambda= 2.17 \mu$m) and Paschen $\beta$ (Pa$\beta$, $\lambda = 1.28 \mu$m) hydrogen emission lines for a single epoch (2003/05/23 UT, see Table \ref{table:observations}). We were able to determine a Br$\gamma$ equivalent width for all four stars in our sample, and a Pa$\beta$ equivalent width for three out of the four stars. We were not able to measure the Pa$\beta$ equivalent width for VV CrA B because VV CrA B was too faint in the J band to be observed simultaneously with VV CrA A. Because we were not able to correct our J and K band spectra for veiling, these EWs are likely only lower limits (Table \ref{table:acclum}). We used these equivalent width lower limits to estimate the minimum accretion luminosity in both the Pa$\beta$ and Br$\gamma$ emission lines; our measurements are presented below along with the discussion of our calculations.

\subsection{Accretion Luminosity}

\begin{deluxetable*}{cccCCCCCC}
\tablewidth{0pt} 
\tablecaption{Lower Limits for J and K Band EWs, Accretion Luminosities, and $\dot{M}$ \label{table:acclum}}
\tablehead{
\colhead{Star} &  \colhead{$R_{\star}$} & \colhead{Line} & \colhead{EW} & \colhead{Total flux} & \colhead{$\log(\frac{L_{line}}{L_{\odot}})$} & \colhead{$\log (\frac{L_{acc}}{L_{\odot}})$} & \colhead{$\log(\frac{L_{acc}}{L_{\star}})$} & \colhead{$\dot{M}$}\\
\colhead{} & \colhead{($R_{\odot}$)} & \colhead{} & \colhead{(\AA)} & \colhead{(Jy)} & \colhead{} & \colhead{} & \colhead{} & \colhead{($\frac{M_{\odot}}{yr}$)}
} 
\startdata 
\multirow{2}{*}{S CrA A} & \multirow{2}{*}{1.66} & Pa$\beta$ & 22 & 1 & -2.68 & -0.08 & 0.1 & 7.8 \times 10^{-8}\\
	& & Br$\gamma$ & 7.3 & 2.2 & -3.27 & 0.13 & 0.31 & 1.28 \times 10^{-7}\\
	\hline
\multirow{2}{*}{S CrA B} & \multirow{2}{*}{1.41} & Pa$\beta$ & 20 & 0.43 & -3.08 & -0.51 & -0.06 & 3.84 \times 10^{-8}\\
	& & Br$\gamma$ & 5.5 & 0.71 & -3.87 & -0.59 & -0.14 & 3.22 \times 10^{-8}\\
	\hline
\multirow{2}{*}{VV CrA } & \multirow{2}{*}{1.52} & Pa$\beta$ & 14 & 0.28 & -3.09 & -0.52 & -0.19 & 3.33 \times 10^{-8}\\
	& & Br$\gamma$ & 30 & 1.66 & -3.12 & 0.31 & 0.64 & 2.26 \times 10^{-8}\\
	\hline
\multirow{2}{*}{VV CrA B} & \multirow{2}{*}{1.41} & Pa$\beta$ & \nodata & 0.04 & \nodata & \nodata & \nodata & \nodata\\
	& & Br$\gamma$ & 4 & 2.02 & -3.56 & -0.21 & 0.24 & 7.73 \times 10^{-8}\\
\enddata
\tablecomments{Fluxes are from the combination of 2MASS/WISE data and our flux ratios, and are corrected for extinction. Pa$\beta$ fluxes are derived from J band and Br$\gamma$ fluxes are derived from K band. We assume $L_{\odot} = 3.848 \times 10^{26}$ W.}
\end{deluxetable*}

We used 2MASS (JHK-band) and WISE (L-band) photometry (Table \ref{table:mags}) and our flux ratios (Table \ref{table:flux_ratios}) to determine the flux for each star at JHKL (Table \ref{table:SEDs}). We then combined our Br$\gamma$ and Pa$\beta$ equivalent width measurements with those fluxes to estimate line luminosities. Following a linear relation derived by \citet{2017A&A...600A..20A} from a large sample of class II and transitional disk objects, we determined the Br$\gamma$ and Pa$\beta$ accretion luminosities for S CrA A and B and VV CrA A, and the Br$\gamma$ accretion luminosity for VV CrA B. The results are shown in Table \ref{table:acclum}. We did not correct the Pa$\beta$ and Br$\gamma$ line EWs by removing a continuum veiling component. Therefore, the EW values in Table 6, and correspondingly the derived line luminosities and accretion luminosities are lower limits. In addition, the photometry, flux ratios, and equivalent widths used to derive the accretion luminosity are from different epochs, 10-15 years apart. Thus, we expect some additional uncertainty \citep[$\sim 1$ mag;][]{2008A&A...482L..35G} in the derived stellar luminosity as a result of this mismatch. However, since T Tauri stars are not typically more variable than 1 magnitude, we do not expect the luminosity to be so far off as to make our conclusions inaccurate.

After correcting our fluxes for extinction using the $A_{V}$ found in \citet{2018A&A...617A..83V} for each star and shown in Table \ref{table:indices}, and defining $A_{\lambda} = A_{v}[0.55/\lambda (\mu m)]^{1.6}$ \citep{2003ApJ...584..853P}, we find line luminosities of $-3.9 < \log(\frac{L_{line}}{L_{\odot}}) < -2.7$, and accretion luminosities of $-0.6 < \log(\frac{L_{acc}}{L_{\odot}}) < 0.3$,  Referring to the plots shown in Figure E.1 of \citet{2017A&A...600A..20A}, those authors found that their sample of Lupus YSOs, with spectral types mostly ranging from K7 to M7, have line and accretion luminosity ranges of approximately $-7 < \log(\frac{L_{line}}{L_{\odot}}) < -3$ and $-5 < \log(\frac{L_{acc}}{L_{\odot}}) < 0$, respectively. S CrA and VV CrA thus have relatively high accretion luminosities compared to \citet{2017A&A...600A..20A}'s $\sim$ 3 Myr old Lupus sample. However, several objects in the Class II phase have been observed to have $L_{acc}/L_{\star} > 1$ \citep{1998ApJ...509..802C, 2006A&A...452..245N, 2017A&A...604A.127M}. Thus the high accretion luminosity values relative to the\citet{2017A&A...600A..20A} sample may be characteristic of the relatively early evolutionary stage of the Corona Australis region. 

{To calculate $L_{acc}/L_{\star}$, we first estimated the stellar luminosity using the best fit mass and age from the \citet{baraffe} isochrones, which include a predicted luminosity for each mass. Then, following \citet{2008ApJ...681..594H}, we assume that 

\begin{equation*}
\dot{M} = (1 - \frac{R_{\star}}{R_{in}})^{-1} \frac{L_{acc}R_{\star}}{GM_{\star}}
\end{equation*}

and that $R_{in} = 5 R_{\star}$ \citep{1998ApJ...492..323G}. With the spectral type to $T_{eff}$ conversion in \citet{2003ApJ...590..348L}, we converted our spectral types to $T_{eff}$. We derived $R_{\star}$ and $M_{\star}$ for our target stars from the \citet{baraffe} models, and list them in Tables \ref{table:acclum} and \ref{table:indices}, respectively. Using those stellar parameters, we derive mass accretion rates, shown in Table \ref{table:acclum}. We find the mass accretion rate for S CrA A to be $\sim 9 \times 10^{-8} M_{\odot}/yr$, the mass accretion rate for S CrA B and VV CrA A to be $\sim 3 \times 10^{-8} M_{\odot}/yr$, and that of VV CrA B to be $\sim 8 \times 10^{-8} M_{\odot}/yr$. The derived accretion rate for VV CrA A agrees with the mass accretion rate estimate from \citet{2016MNRAS.458.2476S} of $\sim 4 \times 10^{-8} M_{\odot}/yr$, but we find that the accretion rates for S CrA A and B are lower than those measured in 2018 by \citet{2018A&A...614A.117G}, who found that the accretion rate was $\sim 5 \times 10^{-7} M_{\odot}/yr$, about an order of magnitude higher than we measure for both stars.

\subsection{Veiling/Equivalent Width correlation}
\begin{figure}
\includegraphics[width = \linewidth, keepaspectratio = true]{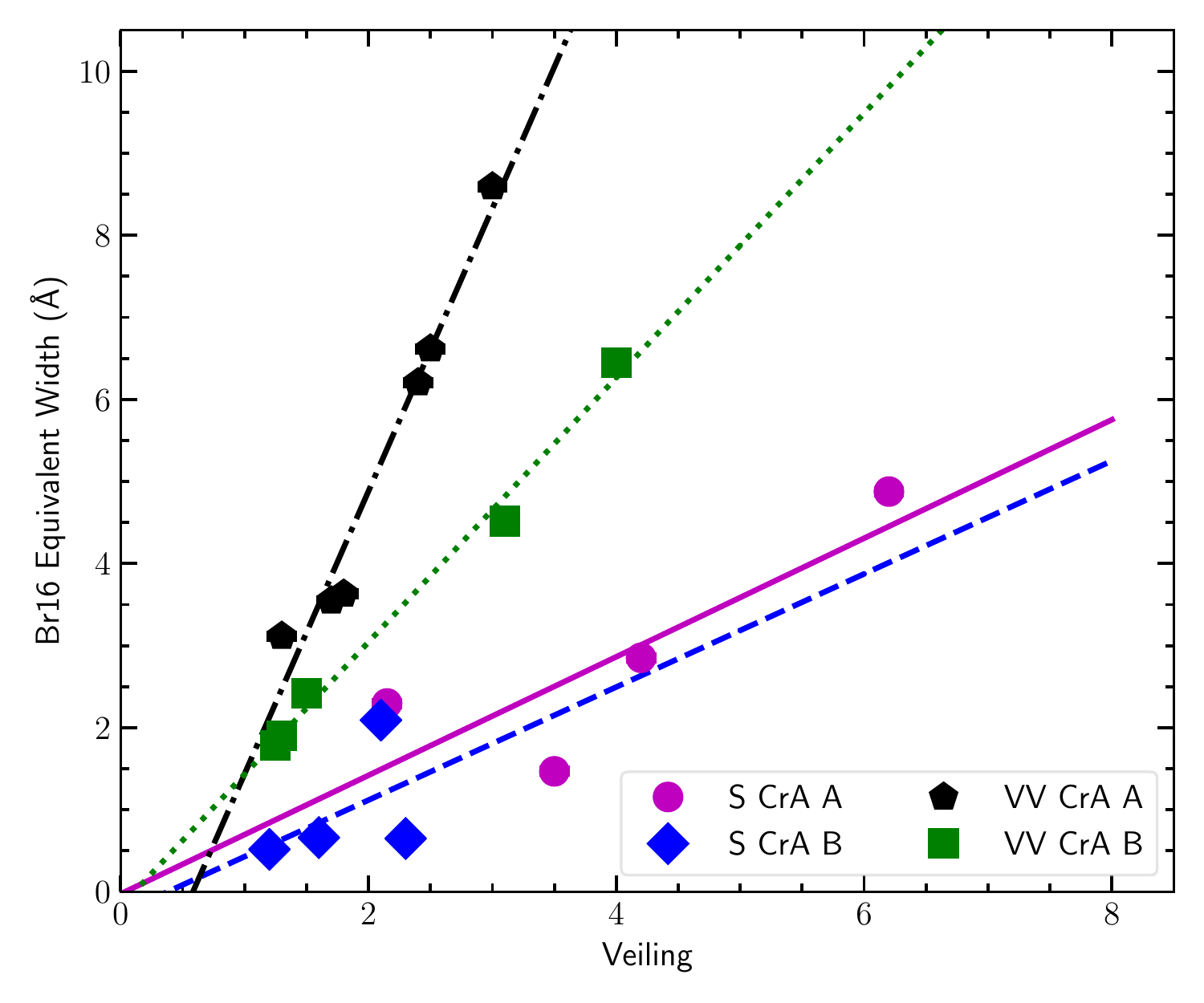}
 \caption{Veiling-corrected Br16 equivalent width versus continuum veiling. Uncertainties are on the order of or smaller than the plotted points.} 
\label{fig:veiling_corr}
\end{figure}

\begin{deluxetable}{cCcc}
\tablewidth{0pt} 
\tablecaption{Linear regression results \label{table:reg_res}}
\tablehead{
\colhead{Star} & \colhead{Slope} & \colhead{Intercept} & \colhead{P-value} 
} 
\startdata 
S CrA A & 0.7 \pm 0.3 & -0.02 & 0.16\\
S CrA B & 0.7 \pm 0.9 & -0.26 & 0.54\\
VV CrA A & 3.4 \pm 0.3 & -2.01 & 0.0005\\
VV CrA B & 1.61 \pm 0.09 & -0.18 & 0.0004\\
\enddata
\end{deluxetable}

We plotted the veiling-corrected Br16 equivalent widths against the H band continuum veiling for each star at each epoch (Figure \ref{fig:veiling_corr}). We performed a linear regression fit to the data; a P-value of less than 0.05 indicates a statistically significant correlation (Table \ref{table:reg_res}). We found that for VV CrA and B the Br16 equivalent width shows a strong linear correlation with the veiling; this is not as clearly the case for S CrA. The Br 16 equivalent widths are on average larger for the components of VV CrA. This suggests that the correlation with veiling may be associated with more accretion. Of the stars in the systems studied, S CrA A has the largest range of veiling variations, $\sim$2-6, and S CrA B has the smallest range, $\sim$1-2.

\subsection{SEDs}

We used unresolved JHKL photometry (Table \ref{table:mags}) with our flux ratios (Table \ref{table:flux_ratios}) to derive NIR stellar fluxes for individual components of S CrA and VV CrA. We show the component-resolved K-band ($2.2 \mu$m) through Q-band ($\sim 18 \mu$m) photometry of \citet{2006ApJ...636..932M} and \citet{2016MNRAS.458.2476S} for S CrA and VV CrA, respectively, in Table \ref{table:SEDs}. Our SEDs are shown in Figure \ref{fig:SEDs}. A representative spectrum for each star from \citet{0067-0049-207-1-5} was normalized at J band, where extinction is minimal, and is overplotted in each panel of Figure \ref{fig:SEDs}. The \citet{0067-0049-207-1-5} SEDs are derived from a sample of YSOs in Lynds 1641, which have an approximate age of $\sim$ 1 Myr, and span a mass range from 0.1 to 3 M$_{\odot}$. Each median SED is constructed for the measured spectral type of each CrA star.

\begin{deluxetable*}{CCCCCC}
\tablecaption{NIR and MIR photometry used for SEDs \label{table:SEDs}}
\tablecolumns{6}
\tablewidth{0pt}
\tablehead{
\colhead{Wavelength} & \colhead{S CrA A} & \colhead{S CrA B} & \colhead{Wavelength} & \colhead{VV CrA A} & \colhead{VV CrA B}\\
\colhead{$\mu$m} & \colhead{(Jy)} & \colhead{(Jy)} & \colhead{$\mu$m} & \colhead{(Jy)} & \colhead{(Jy)}
}
\startdata
1.235\tablenotemark{a} & $0.61 \pm 0.02$ & $0.30 \pm 0.02$ & 1.235\tablenotemark{a} & $ 0.025\pm0.005 $ & $ 0.24\pm0.01 $\\
1.235 & $0.590 \pm 0.006$ & $0.25 \pm 0.03$ & 1.235 & $ 0.1787\pm0.0001 $ & $ 0.002\pm0.005 $\\
1.66\tablenotemark{a}  & $0.98 \pm 0.04$ & $0.48 \pm 0.04$ & 1.66\tablenotemark{a} & $ 0.13\pm0.01 $ & $ 0.44\pm0.02 $\\
1.66 & $1.13 \pm 0.02$ & $0.42 \pm 0.05$& 1.66 & $ 0.669\pm0.004 $ & $ 0.07\pm0.03 $\\
2.16\tablenotemark{a}  & $1.47 \pm 0.08$ & $0.77 \pm 0.06$ & 2.16\tablenotemark{a} & $ 0.73\pm0.03 $ & $ 0.77\pm0.04 $\\
2.16 & $1.80 \pm 0.02$ & $0.60 \pm 0.07$ & 2.16 & $ 1.40\pm0.02 $ & $ 0.73\pm0.08 $\\
2.21\tablenotemark{b} & $ 2.4 \pm 0.2 $ & $ 0.73 \pm 0.09 $ & 2.16\tablenotemark{c} & $ 1.63 \pm 0.23 $ & $ 0.846 \pm 0.23 $\\
3.45\tablenotemark{a}  & $2.64 \pm 0$ & $1.05 \pm 0$ & 3.45\tablenotemark{a} & $ 9.3\pm0.3 $ & $ 0.86\pm0.03 $\\
3.45 & $2.11 \pm 0.07$ & $0.72 \pm 0.4$ & 3.45 & $ 2.7\pm0.6 $ & $ 2.7\pm0.9 $\\
3.5 \tablenotemark{b}& $ 4.0 \pm 0.2 $ & $ 1.15 \pm 0.03 $ & 3.8\tablenotemark{c} & $ 1.88 \pm 0.35 $ & $ 1.69 \pm 0.35 $ \\
\nodata & \nodata & \nodata & 4.8\tablenotemark{c} & $ 3.5 \pm 0.65 $ & $ 2.6 \pm 0.65 $ \\
\nodata & \nodata & \nodata & 7.73\tablenotemark{c} & $ 13.7 \pm 0.7 $ & $ 7 \pm 0.7 $ \\
8.8\tablenotemark{b} & $ 1.9 \pm 0.2 $ & $ 1.4 \pm 0.1 $ & 8.74\tablenotemark{c} & $ 14.4 \pm 0.2 $ & $ 4.5 \pm 0.2 $ \\
10.8\tablenotemark{b} & $ 3.0 \pm 0.2 $ & $ 1.20 \pm 0.06 $ & 10.35\tablenotemark{c} & $ 15 \pm 0.3 $ & $ 3.3 \pm 0.3 $ \\
12.5\tablenotemark{b} & $ 1.8 \pm 0.1 $ & $ 1.34 \pm 0.09 $ & 12.33\tablenotemark{c} & $ 20.5 \pm 0.7 $ & $ 8.7 \pm 0.7 $\\
18.1\tablenotemark{b} & $ 4.3 \pm 0.4 $ & $2.39 \pm 0.3 $ & 18.3\tablenotemark{c} & $ 28.4 \pm 1.6 $ & $ 11.1 \pm 1.6 $ \\
\enddata
\tablenotetext{a}{Data from \cite{2003ApJ...584..853P}.}
\tablenotetext{b}{Data from \cite{2006ApJ...636..932M}.}
\tablenotetext{c}{Data from \cite{2016MNRAS.458.2476S}.}

\tablecomments{Data not marked with a note are from this work.}
\end{deluxetable*}

\begin{figure*}
\includegraphics[width = \linewidth, keepaspectratio = true]{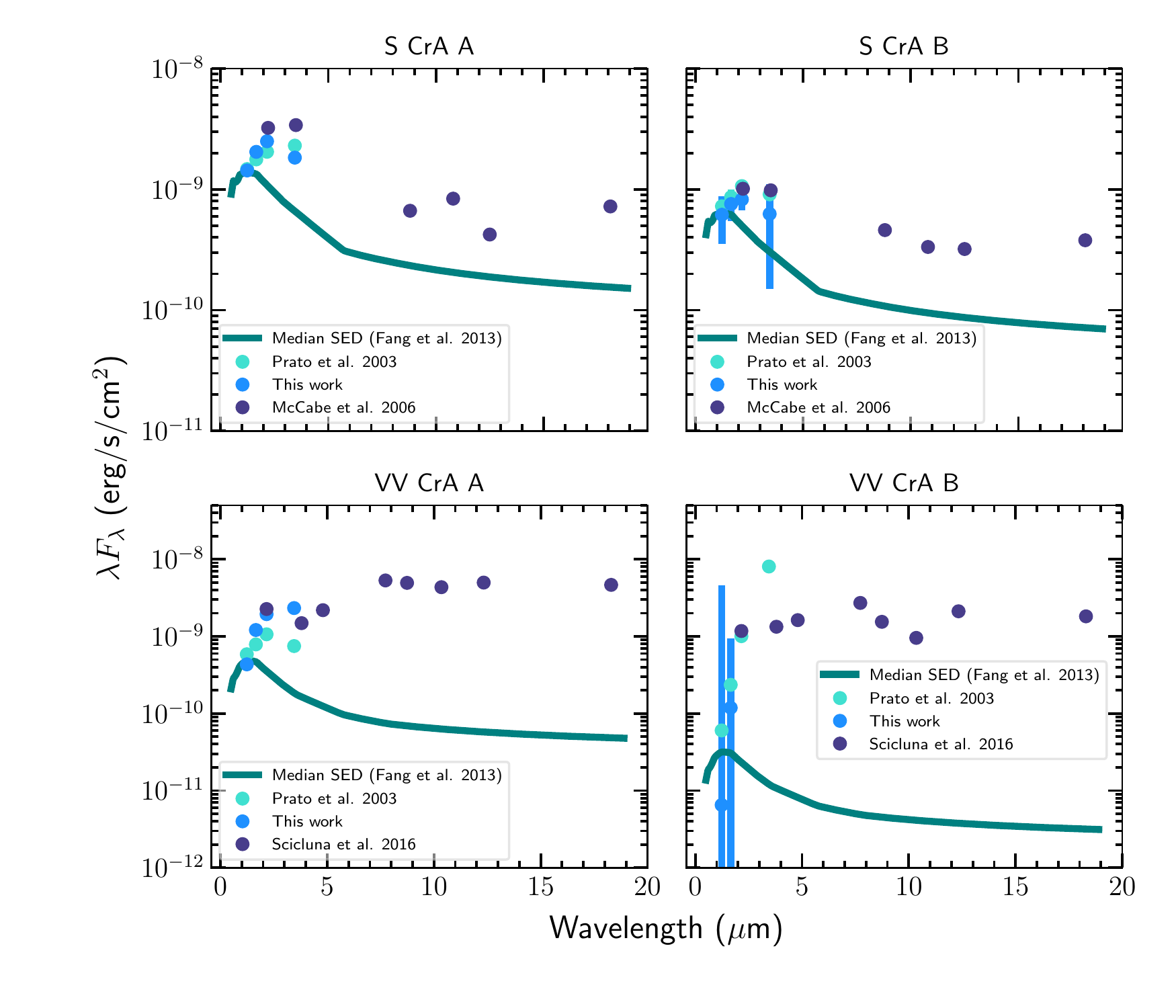}
\caption{Spectral energy distributions for S CrA and VV CrA. Median SEDs of young stars of the same spectral type as S CrA and VV CrA are plotted as the turquoise line \citep{0067-0049-207-1-5}. The MIR data are from \citet{2006ApJ...636..932M} and \citet{2016MNRAS.458.2476S} for S CrA and VV CrA, respectively, and are summarized in Table \ref{table:SEDs}. Error bars are typically smaller than the points.} 
\label{fig:SEDs}
\end{figure*}

 A protostar's class is typically identified using the spectral index of its SED, calculated between 2.2 and 10 $\mu$m. T Tauri stars are Class II objects, defined as having a spectral index between -0.3 and -1.6 \citep{1987IAUS..115....1L, 1994ApJ...434..614G}. Spectral indices between 0.3 and -0.3 describe flat spectrum sources, which have either a flat or double-peaked SED. Spectral indices greater than 0.3 typically indicate a Class I source, i.e. an obscured YSO consisting of a very young star, an optically thick disk, and an infalling envelope \citep{1987ApJ...312..788A, 2001ApJ...551..357W}. 

\citet{1994ApJ...434..614G} define the spectral index as 
\begin{equation*}
\alpha = \frac{d \log(\lambda F_{\lambda})}{d \log(\lambda)} = \alpha_{\lambda_{2}} - \alpha_{\lambda_{1}}
\end{equation*}

We first corrected our fluxes for extinction as described in Section 3.5, then calculated the spectral indices for S CrA and VV CrA using the equation above for $\lambda_{1} = N \approx 10.5 \mu m$ and $\lambda_{2} = K \approx 2.3 \mu m$. We find that the spectral indices for both S CrA A and B indicate that they are flat spectrum objects, rather than Class II objects. The spectral indices for VV CrA A and B are 0.13 and 0.37, respectively, indicating that VV CrA A is a flat spectrum source. Although the SED for VV CrA B appears relatively flat, the spectral index indicates that it is a Class I source. We note an absorption feature at $\sim 10 \mu$m in VV CrA A and B, which we interpret as a silicate absorption feature. The silicate emission in S CrA A agrees with its spectral index to verify its classification as a Class II source, while the silicate absorption for VV CrA A and B corresponds with their identification via spectral index as Class 
I or slightly later sources. These results also agree with \citet{2013A&A...551A..34S}, who note the silicate emission in S CrA.

\section{Discussion}\label{disc}
Our results suggest that VV CrA and S CrA are examples of systems in transition between the Class I and Class II stages of stellar evolution. Figure \ref{fig:veiling_corr} indicates that our understanding of the relationship between NIR veiling and accretion is likely incomplete: for the stars in VV CrA is there a strong correlation between Br 16 equivalent width and NIR veiling. The contrasting weaker correlation for the stars in S CrA points to the possibility that there may be multiple physical processes producing NIR veiling which have different degrees of importance in different systems. The similarity in the correlation between the two stars in each system (which is stronger in VV CrA) indicates that this result may be attributed to systemic properties (e.g., age, environment) rather than individual stellar properties such as mass, T$_{eff}$, etc.. However, the similarity in the stellar properties of the four stars in these systems, coupled with our relative lack of knowledge about their environments, limits the strength of any conclusions we can draw.

In young stars with active disks, the accretion shock produces the optical veiling \citep{1998ApJ...509..802C, 2008ApJ...681..594H}. To account for NIR veiling, models have implicated an origin in the emission of heated grains at the dust sublimation radius of an inner disk wall, material surrounding hot accretion spots, and warm gas inside of the dust sublimation radius \citep[e.g.,][]{2001A&A...371..186N, 2003ApJ...597L.149M, 2011ApJ...730...73F}. \citet{1999A&A...352..517F} argued that the amount of NIR veiling observed in CTTS is too large to originate from a single mechanism and posit that there are multiple components to the NIR veiling. Because of the strong accretion in our CrA targets, which causes Br 16 emission in all components, our H-band observations provide the opportunity to simultaneously measure both the NIR veiling and the accretion intensity. This circumvents the uncertainties inherent in non-simultaneous observations and underscores the differences in the accretion/disk properties for each star plotted in Figure \ref{fig:veiling_corr}. While VV CrA A has the greatest Br 16 equivalent width, the stars in S CrA are more heavily veiled. The IRC, VV CrA B, lies in between these extremes, which, given its early evolutionary stage, is unexpected.

For each of the four stars in these systems, the NIR veiling appears to trace the accretion with different proportionality. How the inner disk rim model of \citet{2003ApJ...597L.149M} and others might account for this is unclear. One possibility is that the slope of the EW versus NIR veiling relation is a function of the dust truncation radius, R$_{d}$, and may indicate whether R$_{d}$ is at or further out than the corotation radius from which gas accretion originates. Further work should examine this relation in a larger sample of young stars at early evolutionary stages with a variety of observational cadences to decompose the various components contributing to NIR veiling. The fraction of NIR veiling that originates in an inner disk wall compared to the accretion shock may be a sensitive function of age; although the ages of S CrA and VV CrA are similar, the stars in the latter system could be in a slightly earlier evolutionary stage.
 
We have measured the spectral types of these four stars accurately for the first time. These observations have historically been challenging given the need for a combination of high spectral resolution and a high signal to noise ratio. However, a relative lack of deep photospheric absorption lines even in our spectra has restricted our ability to determine the spectral type to better than 1 spectral class. Multiple epochs of observations offset this limitation by providing a cross-check of our initial spectral type determination over at least four epochs for each star. 

We find that the estimated ages of S CrA and VV CrA differ by $\sim$1 Myr. Although the stars in VV CrA appear to be at an earlier evolutionary stage than those in S CrA, they fall on an older isochrone. There are many examples of CTTS and WTTS in large populations with an overlapping range of ages \citep[e.g.,][]{1998AJ....116.1816H}, thus this result is not necessarily inconsistent with S CrA and VV CrA both being coeval. Furthermore, ages derived from isochrones are uncertain \citep[e.g.,][]{2014prpl.conf..219S}. The difference in ages falls within our $1 \sigma$ error bars for the isochrone ages of about $\sim$ 1 Myr \citep{2014prpl.conf..219S}. Our assumption of idential distances to S CrA and VV CrA may also be inaccurate.

The presence of high amplitude and variable NIR veiling and Br16 emission implies a high accretion rate onto a star that is surrounded by a substantial circumstellar disk. For S CrA this is verified by \citet{2019arXiv190402409C}, who find that S CrA A and B have disk masses of 95 and 102 $M_{\bigoplus}$, respectively, which is much higher than the average disk mass in CrA or in other star forming regions \citep{2019arXiv190402409C}. Most CTTSs are not as obscured by veiling as S CrA and VV CrA, suggesting that these two systems are either younger than CTTSs or are early in the CTTS stage. The SED analysis we performed substantiates this by showing that the components of VV CrA have a flatter spectral slope and silicate absorption, rather than emission, suggesting that they are surrounded by more cold material and dust than S CrA and hence are in an earlier evolutionary stage. 

Although there are many factors that may influence SED spectral slope besides age, a large IR excess is typically interpreted as a more embedded, and thus less evolved, source \citep{1994ApJ...434..614G}. This conclusion is substantiated by our measured spectral types and relative H-band veiling values, which all indicate generally young sources, with VV CrA B being the least evolved and the two stars in S CrA being more evolved. Our CrA stars have accretion luminosities and IR excesses (Figure \ref{fig:SEDs}) much larger than the \citet{0067-0049-207-1-5} median SEDs for class II stars, suggesting relative youth. Furthermore, the NIR spectra of our targets (Figures \ref{fig:all spectra} and \ref{fig:JK spectra}) look markedly different from most CTTS (e.g., spectra available at \url{http://jumar.lowell.edu/BinaryStars/}, Prato et al. in prep). This may be an indication of a phase of relatively rapid changes in qualitative appearance resulting from relatively small quantitative changes in accretion and NIR veiling, caused by increased activity explained by a younger age.

Observations at multiple epochs are essential for understanding young stars themselves. In these two systems alone, our time-domain spectral data have demonstrated the presence of variable NIR veiling of unknown origin, and have allowed us to determine fundamental parameters such as spectral type much more definitively than is typically possible for stars of this age and degree of variability and NIR veiling. We cannot understand the interplay between young stars and their environments without a clearer picture of the relationship between circumstellar disks and stars. This is crucial for developing more nuanced models for disk evolution and planet formation in multiple systems, in which most stars are located. Since multiplicity potentially shapes the evolution of circumstellar disks \citep[e.g.,][]{2016AJ....152....8K}, time-domain, multi-wavelength studies that complement bulk investigations of statistical populations of young binaries are vital for a deeper understanding of star, disk, and planet evolution.

\section{Summary}\label{sum}
We have characterized each star in the young binary systems S CrA and VV CrA. We measured the H band veiling, determined spectral types, and analyzed three emission lines in the NIR. No evidence of higher-order multiplicity was detected. We have constructed SEDs from a combination of archival and new data, and estimated the evolutionary stage of each star based on the spectral indices of those SEDs. We have calculated the accretion luminosity using Br$\gamma$ and Pa$\beta$, and find it to be higher than is typical for most T Tauri stars, suggesting that these systems are relatively young. This is consistent with the Class I characterization for the IRC VV CrA B, first proposed in the mid-1990s \citep{1997ApJ...480..741K}, and an intermediate stage bridging the Class I and Class II phases for the other components, as opposed to the classical Class II categorization of VV CrA A and S CrA A and B. The stars in S CrA and VV CrA all have approximately the same mass, age, and temperature. VV CrA B, typically characterized as the secondary star in the system, has an earlier spectral type than VV CrA A and is therefore the more massive component, although it is fainter at shorter wavelengths given its greater obscuration.

Time-domain NIR, high-resolution spectroscopy provides a powerful tool for the study of young binary stars and their environments. The ability to simultaneously link stellar and circumstellar environmental properties over multiple epochs provides greater insight and a deeper understanding of binary star and circumstellar disk evolution. Working in the NIR facilitates observations of relatively obscured systems such as S CrA and VV CrA, high-resolution spectroscopy permits detailed measurements of stellar and accretion properties, and time-domain observations reveal the extent of variability in the dynamic circumstellar environments surrounding these young stars. By comparing the detailed stellar and environmental properties of the component stars in young binaries, which presumably share the same environments and similar evolutionary paths, we can distinguish between characteristics resulting from physical differences in the stars themselves, e.g., mass, radius, luminosity, random processes, and initial conditions which may have been in place since the early deeply embedded Class 0 phase of a star's evolution.

\section*{Acknowledgments}
We thank the anonymous referee for their helpful suggestions and contributions. We thank the Keck Observatory OAs and staff for their technical expertise and we acknowledge helpful discussions with Christopher Johns-Krull and Sean Graham. We also thank Min Fang for sharing his median TTS SEDs with us. This research was supported in part by NSF grants AST-1313399 (to L. P.) and AST-1461200. G.H.S. acknowledges support from a NASA Keck PI Data Award administered through NExScI. 
This work has made use of data from the European Space Agency (ESA) mission
{\it Gaia} (\url{https://www.cosmos.esa.int/gaia}), processed by the {\it Gaia}
Data Processing and Analysis Consortium (DPAC,
\url{https://www.cosmos.esa.int/web/gaia/dpac/consortium}). Funding for the DPAC
has been provided by national institutions, in particular the institutions
participating in the {\it Gaia} Multilateral Agreement. This research has made use of the Keck Observatory Archive (KOA), which is operated by the W. M. Keck Observatory and the NASA Exoplanet Science Institute (NExScI), under contract with the National Aeronautics and Space Administration. Some of the data presented herein were obtained at the W. M. Keck Observatory, which is operated as a scientific partnership among the California Institute of Technology, the University of California and the National Aeronautics and Space Administration. The Observatory was made possible by the generous financial support of the W. M. Keck Foundation. The authors wish to recognize and acknowledge the very significant cultural role and reverence that the summit of Mauna Kea has always had within the Indigenous Hawaiian community. We are most fortunate to have the opportunity to conduct observations from this mountain. 

\facility{Keck:II (ESI)}

\bibliography{SVV_paper.bib}

\end{document}